\documentclass[preprint,amsmath,amssymb]{revtex4}

\usepackage{graphicx}
\usepackage{dcolumn}
\usepackage{bm}

\begin{document}

\title{Precise study of the Efimov three-particle spectrum \\
and structure functions within variational approach}

\author{B. E. Grinyuk, M. V. Kuzmenko, I. V. Simenog}

\affiliation{Bogolyubov Institute for Theoretical Physics of the
NAS of Ukraine, \\ Metrolohichna Str., 14b, Kyiv-143, 03143,
Ukraine}

\date{\today}

\begin{abstract}
A precise study within variational approach of the basic
properties of the three-particle spectrum and structure functions
with Gaussian potential near the critical coupling constant of
interaction where the Efimov effect takes place is carried out. A
method is developed to calculate highly excited states with very
small binding energies, and numerical analysis is carried out for
the ground and three excited energy states. For these states,
one-particle density distributions, formfactors, pair correlation
functions, and momentum distributions are calculated. It is found
that the second excited state has already all the basic features
of a level from the infinite Efimov series. An essential asymmetry
is found in the position of energy levels with respect to the
critical constant point. A halo-type structure is revealed in the
one-particle density distributions, and formfactors are shown to
have specific dips of finite depth, the number of dips being equal
to the number of the state. The behaviour of pair correlation
functions and momentum distributions is studied for three-particle
states.
\end{abstract}

\maketitle

\section{Introduction}
As known, the Efimov effect~\cite{R1,R2,R3} reveals itself in a
system of three particles with finite-range interaction as the
appearance of the infinite number of levels near the threshold
with very small energies under the condition that the two-particle
subsystems have an infinitely small binding energy or an
infinitely large scattering length. In the case of three identical
particles, the energy ratio for neighboring highly excited
near-threshold Efimov states~\cite{R1,R4} is $\lambda _E =
\lim\limits_{n \to \infty } (E_n / E_{n + 1} ) \approx 515$. In
three-nucleon systems, the conditions necessary for the effect to
take place can be easily broken by taking into account the spin
structure of the nuclear interaction~\cite{R3,R6} or the Coulomb
long-range repulsive potential.

Theoretical studies of the Efimov near-threshold energy spectrum
were carried out both by asymptotic expansions with the use of the
small parameter $r_0 / a \ll 1$  (being the ratio of the range of
forces to the two-particle scattering
length)~\cite{R1,R2,R3,R5,R6} and by numerical
calculations~\cite{R7,R8} based on the Faddeev equations by using
separable interaction potentials. The main properties of the
energy states are universal to the great extent and do not depend
on a specific form of the potential. For the special choice of an
interaction potential in the form of two components with
essentially different radii, new additional excited energy levels
appear in the three-particle system --- the ``trap''
effect~\cite{R9}. In the case of local potentials, numerical
calculations of the Efimov energy spectrum were not carried out
systematically because of essential difficulties due to the
different scales of very small near-threshold excited states
energies and the ground-state energy. This requires to develop the
precise methods of calculation of the highly excited weakly
bounded states of three-particle systems with local interactions.

Due to the use of Gaussian bases in the variational calculations
of a few-particle systems~\cite{R10}, noticeable progress was
achieved in the calculations of the main properties of few-body
systems~\cite{R11,R12,R13,R14} with interactions of different
nature in recent years. In the present paper, within the precise
variational approach, a study of the near-threshold Efimov states
of a three-particle system with Gaussian potential is carried out
up to the third excited level, and all the main structure
functions for various energy states are found.

\section{Statement of the Problem}
Consider a system of three identical particles
\begin{equation}
H =  - \frac{\hbar ^2}{2m}\sum\limits_{i = 1}^3 \Delta _i - V_0
\sum\limits_{i > j = 1}^3 \exp \left( - \frac{r_{ij}^2 }{r_0^2
}\right) \label{E1}
\end{equation}
with a two-particle interaction taken in the Gaussian form for
simplicity, where $V_0$ is the intensity and
$r_0$ is the radius of forces. Let us study the properties of
the energy spectrum and the main structure
functions of the system in symmetric (with respect to
permutations of particles) states with zero total angular
momentum in the range of parameters where the Efimov effect
takes place. It is convenient to use the
dimensionless variables and to measure all the distances in $r_0$
and energies in $\hbar ^2 /mr_0^2$. In this
case, the dimensionless Hamiltonian
\begin{equation}
H =  - \frac{1}{2}\sum\limits_{i = 1}^3 \Delta _i  -
g\sum\limits_{i > j = 1}^3 \exp \left( - r_{ij}^2 \right)
\label{E2}
\end{equation}
contains only one combination of physical parameters, the coupling
constant $g = mr_0^2 V_0 /\hbar ^2$, which determines the
properties of the spectrum and wave functions of the system.

The solution of the problem on eigenvalues and eigenfunctions for
bound states is carried out in the framework of the variational
Galerkin method in the Gaussian representation with the use of
special optimization schemes (see~\cite{R12,R13,R14}) which
enhance the convergence and ensure a high accuracy of calculations
with the least dimensions of variational bases. Within this
approach, the variational wave function of symmetric states of the
three-particle system with zero total angular momentum can be
presented as the expansion in Gaussian functions depending on
relative distances $r_{ij}  = \left|{\bf r}_i  - {\bf r}_j
\right|$,
\begin{equation} 
\Psi \left( r_{12}, r_{13}, r_{23} \right) = \sum\limits_{k =
1}^{\cal K} {D_k {\bf S}\left. \left| {\psi _k } \right.
\right\rangle } 
\equiv\sum\limits_{k = 1}^{\cal K} {D_k {\bf
S}\exp \left( - a_k r_{12}^2 - b_k r_{13}^2  - c_k r_{23}^2
\right)}\, , \label{E3}
\end{equation}
with a set of variational parameters $a_k$, $b_k$, $c_k$. For the
bound states, the linear coefficients $D_k$  are found in the
variational approach from the system of linear algebraic equations
of a problem on eigenvalues (which is equivalent to the
Schr\"{o}dinger equation in the chosen Gaussian basis):
\begin{equation}
\sum\limits_{n = 1}^{\cal K} D_n \Bigl(\left\langle {\bf S}\psi _k
\left|H\right|{\bf S}\psi _n  \right\rangle  - E\left\langle {\bf
S}\psi _k \left|  {\bf S}\psi _n \right.\right\rangle \Bigr) =
0,\,\,\, k = 1,2,\dots,{\cal K}. \label{E4}
\end{equation}
Note that all the matrix elements in Eq.~(\ref{E4}) for
Hamiltonian~(\ref{E2}) can be calculated in explicit form. The
chosen method of calculation has proved its high efficiency and
accuracy in a number of problems of a few particles with
interactions of different nature. But, in the case of the
near-threshold Efimov spectrum, we are faced with an additional
difficulty due to very small values of the energies of these
levels. Since the ratio of the Efimov neighboring energy levels $
\lambda _E  = \lim \limits_{n \to \infty } (E_n /E_{n + 1} )
\approx 515$ is rather large, there exists a real difficulty for
all the methods (including the method based on the Faddeev
equations~\cite{R9} and variational methods as well) in
calculating the energies with essentially different scales and the
corresponding wave functions with oscillations at essentially
different distances.

In the present work, the main structure properties of the ground
and three excited states of a system of three particles are
calculated with the coupling constants near $g \approx g_{cr}$,
where a two-particle bound state appears and the Efimov effect
reveals itself. The numerical analysis of highly excited
near-threshold Efimov levels appears to be a nontrivial problem
for all the methods including the variational method because of
the necessity to prepare the ``initial'' configurations for the
wave functions of highly excited near-threshold bound states to
start the minimization of energy in the nonlinear variational
parameters $a_k$, $b_k$, $c_k$.

\section{The Problem of Initial Configurations}
An advantage of the variational method lies in the possibility to
calculate (in the approximation of a given number of the basis
functions) the whole spectrum of low-energy bound states for a
given Hamiltonian or to calculate the energy and wave function of
the chosen ground or excited state separately. The further
minimization in the nonlinear parameters $a_k$, $b_k$, $c_k$
enables one to improve the variational estimation for the energy
and to achieve the best result at a fixed dimension of the basis
in~(\ref{E3}). It is clear that if the initial configuration
gives, in fact, the bound (ground or excited) state, we have to
approach the exact solution for this state by minimizing the
nonlinear parameters $a_k$, $b_k$, $c_k$ and expanding the basis
dimension. However, at a finite and even rather large number
${\cal K}$ of the basis functions in~(\ref{E3}) and with a random
set of the nonlinear parameters $a_k$, $b_k$, $c_k$  (we call it a
random ``initial configuration''), nobody can be sure not only in
the fact that the studied energy level is close to the exact
value, but, even more, that it is below the threshold of a
decomposition into subsystems at all. In the latter case, as a
rule, the further minimization in nonlinear parameters should be
terminated on approaching the rated level to the threshold (from
the side of the continuum spectrum), because the wave function can
have nothing in common with a bound state. Such problems reveal
themselves starting from the second excited state which exists in
a narrow interval near $g \approx g_{cr} = 2.684005$ (the critical
constant for the chosen Gaussian potential). There is no such
problem for the ground state of three particles, whose initial
configuration can be simply prepared even with one Gaussian
function at sufficienly large $g$. The mentioned problem is also
easily solved for the first excited state if one takes several
Gaussian functions, among which at least two or three ones depend
on the parameters $a_k$, $b_k$, $c_k$ taken from the ground-state
function, while the rest have to form a cluster
function~\cite{R13}. This enables one to rather simply construct
the necessary configuration (even in the case where the initial
configuration corresponds to an energy slightly exceeding the
two-particle threshold). For the ground and first excited states,
the problem is simplified also due to the fact that these levels
exist at all the interaction constants $g$ greater than certain
critical values (see Table \ref{tab1}). Therefore, it is
sufficient to enhance the attractive potential, to form the
initial configurations at a sufficiently large $g$, and then, by
gradually decreasing $g$ (the evolution in coupling constant), to
achieve the necessary value of $g$  by minimizing the energy in
the parameters $a_k$, $b_k$, $c_k$. We used both variants to form
the initial configurations for the ground and first excited
states.

For the following excited near-threshold Efimov states with
extremely small binding energies existing within a narrow interval
of $g$ near the critical two-particle constant $g_{cr}$, the
problem of initial configurations requires special methods for its
solution. One way is to use a wider class of Hamiltonians, in
particular, those with an additional interaction giving a
possibility to form relatively easily the configurations of highly
excited levels. Then the above interaction is eliminated step by
step (in order to return to the initial problem) with a
simultaneous change in the parameters $a_k$, $b_k$, $c_k$  within
the variational procedure in order to keep the energies of the
formed configurations below the threshold. The Coulomb attractive
interaction seems to be suitable for this purpose giving rise to
the infinite number of levels for the three-particle system. An
additional advantage of this interaction is the fact that all
matrix elements in the Gaussian basis for the Coulomb potential
are known to be calculated in explicit form.

\begin{figure}[ht]
\centering
\includegraphics[width=4.5in]{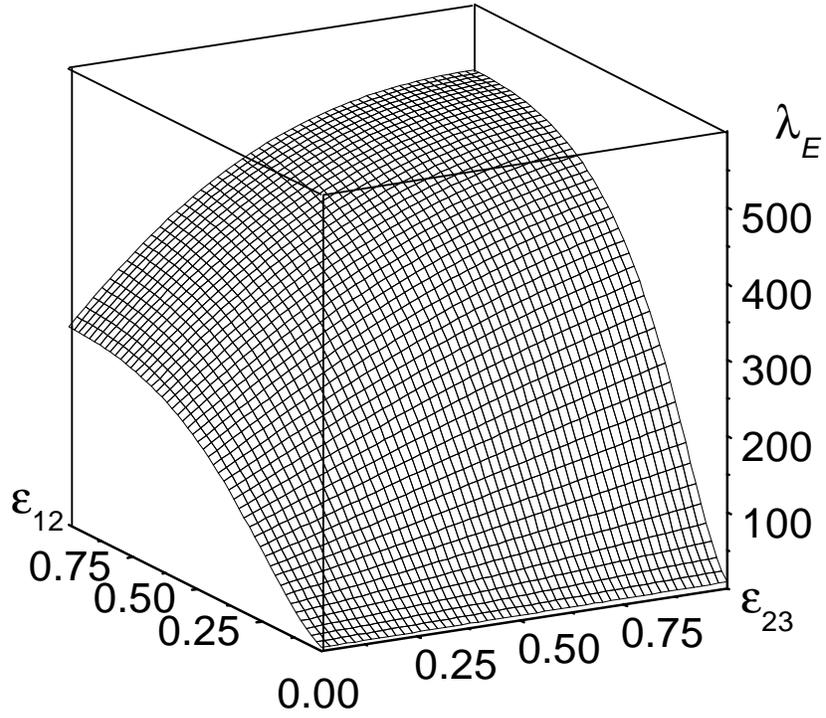}
\caption{Coefficient $\lambda _E$ for the neighboring Efimov
energy levels versus the mass ratio ($\varepsilon _{12}=m_1/m_2$,
and $\varepsilon _{23}=m_2/m_3$) for the system of three
particles.} \label{fig1}
\end{figure}

Another way to prepare the initial configurations is to deal with
a Hamiltonian with different masses of particles. It is known that
the Efimov effect takes place in three-particle systems with
different masses~\cite{R1,R2,R3}. Moreover, the ratio of the
energies of neighboring levels depends significantly on the ratio
of masses (see also~\cite{R15}). From the analysis of the power
asymptotics of a three-particle wave function in the momentum
representation within the approach based on the Faddeev equations,
it can be shown that the following secular equation is the
condition for the solvability of three-particle equations:
\[
L(1;2,3)L(2;1,3) + L(1;3,2)L(3;1,2) + L(2;3,1)L(3;2,1) +
\]
\begin{equation}
L(1;3,2)L(2;1,3)L(3;2,1)+ L(1;2,3)L(2;3,1)L(3;1,2) = 1;
\label{E5}
\end{equation}
where
\begin{eqnarray}
L(i;j,k) \equiv N(\frac{\mu _{jk}}{\mu _i
};\frac{\mu _{jk}}{m_k});\,\,\, N(a;\,b)
\equiv \frac{\sinh (s_0 \theta /2 )}{b \sqrt{a}\, s_0 \,\cosh (s_0
\pi /2 )};\,\,\, \cos \theta \equiv \frac{a - b^2}{a + b^2};
\end{eqnarray}
\begin{eqnarray}
\mu _{jk} \equiv \frac{m_j m_k}{ m_j  + m_k};\,\,\, \mu _i \equiv
\frac{m_i (m_j + m_k )}{ m_i  + m_j + m_k }.
\end{eqnarray}
In the case of equal masses, the secular equation (\ref{E5})
becomes essentially simpler~\cite{R1}:
\begin{equation}
 N\left(\frac{3}{4};\,\frac{1}{2}\right) = \frac{1}{2}\,, \label{E6}
\end{equation}
or
\begin{equation}
\frac{8}{\sqrt 3 } \, \frac{\sinh(\pi s_0/6)}{s_0 \cosh(\pi s_0
/2)} = 1\, , \label{E7}
\end{equation}
and one has for the power $s_0$ of the asymptotics of the Faddeev
equations solution (the Danilov-Minlos-Faddeev-Efimov constant),
$s_0=1.0062378$. This constant determines, in particular, the
ratio of Efimov neighboring energy levels (the scale invariance
multiplier),
\begin{equation}
\lambda _E  = \lim \limits_{n \to \infty } (E_n /E_{n + 1} ) =
\exp \left(\frac{2\pi }{s_0 }\right) \cong 515.035\, . \label{E8}
\end{equation}
Fig.~\ref{fig1} shows the dependence of the ratio $\lambda _E =
\exp (2\pi /s_0)$ of the energies of Efimov neighboring levels on
the ratio of the masses of particles. It is seen clearly that the
case of equal mases is the most unsuitable for the Efimov effect
to reveal itself (the scale multiplier $\lambda _E$ is the
greatest) and, at the same time, it is the most difficult case for
numerical calculations. But, in the case of essentially different
masses, one may have the ratio of energies not too far from 1
instead of (10) (for any ratio of masses, of course, we have
$\lambda _E  > 1$ ). In particular, in the limit $m_1/m_3=m_2/m_3
\to 0$ (the one-center problem), relation (\ref{E5}) yields the
limiting secular equation for $s_0$:
\begin{equation}
\left\{1 + 8 \left[1 + \cosh^{-1}\left(\frac{s_0
\pi}{2}\right)\right]\left[\sinh\left(\frac{s_0
\pi}{4}\right)/s_0\right]^2
\right\}
\cosh^{-2}\left(\frac{s_0 \pi}{2}\right) = 1\, .
\label{E9}
\end{equation}
This results in $s_0  = 1.139759$, and the corresponding ratio of
energies $\lambda _E  = 247.826$. If, vice versa, $m_1/m_3=m_2/m_3
\to \infty$ (the two-center problem), the value of $s_0$ increases
infinitely as $s_0 \to k\sqrt {m_1 / m_3}$, where the coefficient
$k = 0.40103$ is found from the secular equation
\begin{equation}
\frac{\exp ( - k\sqrt {2} )}{k\sqrt {2} } = 1\, . \label{E10}
\end{equation}
The corresponding ratio of energy levels goes to 1 as $ \lambda _E
\to 1 + \frac{2\pi}{k}\sqrt {\frac{m_3}{m_1}}$ at $m_3/m_1 \to 0$.
It should be noted that, due to the essential dependence of the
parameter $\lambda _E$ on the ratio of masses, it is more probable
to observe the Efimov effect in a real physical system just with
different masses of particles whose pairwise interaction is of the
resonance type. Close to such systems might be three-cluster
nuclei consisting of a rather heavy magic core and two neutrons.
In particular, if the nucleus $^{40}$Ca serves to be the core, it
follows from (\ref{E5}) that $\lambda _E \approx 262$ (i.e., it is
twice smaller than that for equal masses). If, to say, we consider
the system of mesoatoms of deuterium and tritium interacting with
an electron, then, due to the small mass of an electron in
comparison with the masses of mesoatoms, one has a large value of
$s_0$ from (\ref{E5}) enough to result in $\lambda _E \cong 1.26$.
This value is not essentially greater than 1 and much less than
that in the case of equal masses. This means that the energies of
neighboring Efimov levels would be of the same order of magnitude
if they were observed in such a system.

Returning to the problem of formation of initial configurations,
we note that the configuration for a given (the $n$-th excited)
state can be prepared rather easily at different masses. Then one
has to restore step by step the equality of masses by minimization
of the energy of this level in nonlinear parameters. In this way,
we can get the necessary variational wave function corresponding
to the configuration of a bound state.

To obtain the initial configurations for variational wave
functions, we paid the main attention to the method of evolution
in the coupling constant $g$ at a fixed (zero) energy, which is
possible because the interaction potential in (\ref{E1}) has a
definite sign. The method, in essence, consists in a variational
procedure with respect to the critical coupling constant $g$ to be
determined for the given ground or excited state at a fixed (zero)
energy. It is the general statement that the variational method
can be formulated with respect to the coupling constant $g$ at a
fixed energy provided that the interaction potential is negative
definite. Taking an arbitrary initial configuration with a
sufficient number of the basis functions with some parameters
$a_k$, $b_k$, $c_k$, we find the corresponding coupling constant
$g$ of the chosen excited state by variational procedure and
advance step by step towards small $g$ by varying the parameters
$a_k$, $b_k$, $c_k$ (or sometimes, if necessary, by spreading the
basis). Thus, we approach the two-particle critical constant from
the upper side, where a two-particle ground state appears. When we
are close enough to the region $g \approx g_{cr}$, where the
Efimov excited state under consideration must exist, we return to
the commonly used variational procedure in energy using the
obtained set of the parameters $a_k$, $b_k$, $c_k$ and start to
increase the parameter $g$. If we were close enough to the
critical constant $g_{cr}$, the three-particle energy level
crosses the two-particle threshold and goes down below the
threshold. Thus, we obtain the required initial configuration.
Such a possibility is connected with the fact that the
two-particle energy threshold behaves itself by the square law
near the critical coupling constant,
\begin{equation}
E_{(2)}  \cong  - \,0.107 \,(g - g_{cr} )^2\, , \label{E11}
\end{equation}
while three-particle levels reveal a linear behavior~\cite{R1},
\begin{equation}
E_n  \cong k_n\,(g - g_{left}^{(n)} ) \, , \label{E12}
\end{equation}
where $g_{left}^{(n)}$ is the critical constant for the $n$-th
level appearance on the left from the critical constant, and $k_n
=\partial E_n/\partial g$ (at the critical point $g =
g_{left}^{(n)}$) is the corresponding angular coefficient (see
Table \ref{tab1}). Using the above method, we succeeded in
preparing the initial configurations for the second and third
excited states which exist near the critical constant $g_{cr}$ and
belong to the Efimov series of levels. Note that the greater the
number of the excited near-threshold Efimov level, the greater are
the difficulties in preparing the initial configuration. It is not
only due to the necessity to increase the dimension of the
variational basis, but also because the nonlinear variational
parameters differ from one another by several orders of magnitude.
Moreover, there appears a considerable quantity of nonlinear
parameters such that they are roughly by $\lambda _E$ times
smaller than those typical of the previous energy level. This
makes it necessary to significantly decrease the steps in the
process of variation in nonlinear parameters and decelerates the
whole procedure of preparing the initial configuration. Because
the energy of already the third excited level is less by nine
orders of magnitude in comparison with the ground state energy and
the dimension of the basis which is necessary to ensure a reliable
accuracy achieves several hundreds of functions, the calculation
of levels higher than the third one becomes a serious problem.

\section{Asymmetry of the Efimov Energy Levels}

\begin{figure}[ht]
\centering \includegraphics[width=4.5in]{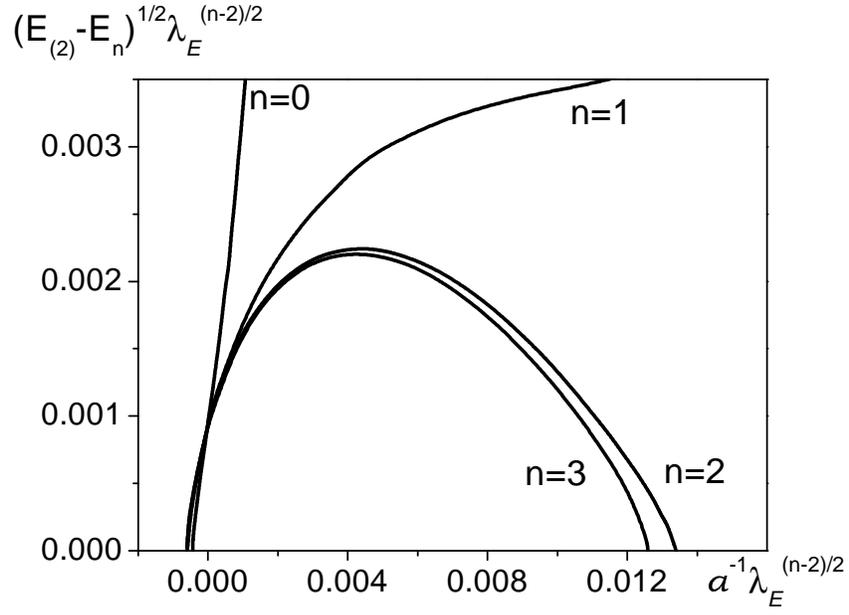}
\caption{Dependence of $(E_{(2)}-E_n)^{1/2}$ on the reciprocal
scattering length $a^{-1}$ for the ground ($n=0$) and excited ($n
= 1, 2, 3$) states. Each of the levels is depicted on another
scale with the use of the multiplier $\lambda _E^{(n-2)/2}$ for
both axes.} \label{fig2}
\end{figure}

\begin{figure}[ht]
\centering \includegraphics[width=4.5in]{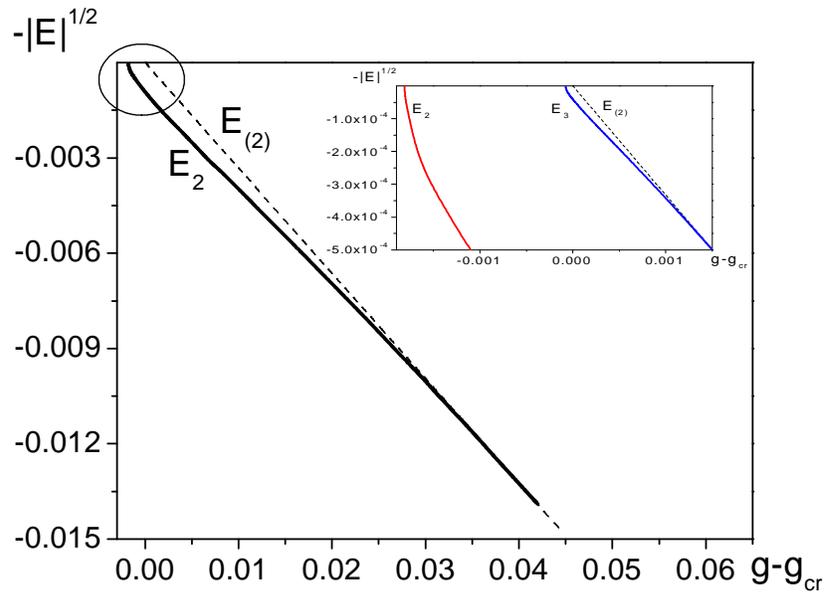}
\caption{Dependence of the second and third excited energy levels
on the coupling constant $g$. The third excited level is shown in
the enlarged fragment.} \label{fig3}
\end{figure}

It is convenient to depict the calculated energies of the ground
and excited states of a three-particle system as the square roots
of the moduli of three-particle energies minus the two-particle
energy threshold versus the reciprocal scattering length (instead
of the coupling constant $g$), because such dependences are
model-independent \cite{R1,R2,R3,R5} to the great extent. Since
the orders of magnitude of the values (of both energies and
reciprocal scattering lengths) are essentially different for
different levels (the ratio for neighboring levels being about
$\sqrt {\lambda _E }  = \exp (\pi / s_0) = 22.692$ ),
Fig.~\ref{fig2} shows each level in its own scale in order to
emphasize the universal properties of highly excited states.
Namely, the horizontal axis depicts $a^{ - 1} \lambda _E^{(n -
2)/2}$, and the vertical axis shows $(E_{(2)}  - E_n )^{1/2}
\lambda _E^{(n - 2)/2}$. The ground and first excited states exist
at any $g$ greater than the corresponding critical values (see
Table \ref{tab1}). The reciprocal scattering length $a^{-1}$ is
approximately proportional to $g - g_{cr}$ near the two-particle
critical constant $g \cong g_{cr}$. Note that the separation of
three-particle levels from the zero energy occurs by law
(\ref{E12}), though this linear behavior takes place only in a
very narrow interval near the corresponding three-particle
critical coupling constant, and the angular coefficient $k_n$
decreases essentially with increase in the number of the excited
state. The second and third excited states appear with increase in
$g$ (in the region $g < g_{cr}$) and then disappear (in the region
$g > g_{cr}$) at the two-particle threshold (the corresponding
values of the coupling constants $g$ and the reciprocal scattering
lengths are given in the second and third columns in
Table \ref{tab1}). A rather good coincidence of the curves for the
second ($n = 2$) and third ($n = 3$) excited levels depicted in
Fig.~\ref{fig2} testifies to that the asymptotic
formulae~\cite{R1,R2} are valid already starting from the second
excited state and indicates the fact that highly excited Efimov
states possess the scaling property. Although we restricted
ourselves by the calculations of only three excited levels, it
should be assumed that the rest of levels is to be determined by
the asymptotic relations, and thus the properties of the whole
spectrum are completely described. A significant asymmetry of the
curves obtained in our calculations for the second and third
excited states relative to the critical concentration point
$g_{cr}$ of the Efimov spectrum testifies to the asymmetry of all
the rest levels. Each of them extends to the right side from the
critical point much farther than to the left. This essential
asymmetry of the levels relative to the point $g = g_{cr}$ (or
$a^{-1} = 0$) is connected with the following fact. Though the
potential of the effective long-range interaction $\sim - R^{-2}$
\cite{R1,R2,R3} ranges up to distances of the order of
$\left|a\right|$, it remains attractive at essentially larger
distances at $a > 0$. Whereas, at $a < 0$, it becomes practically
zero out of the region of about $\left|a\right|$ (see
also~\cite{R15}). The above-mentioned asymmetry is clearly seen in
Fig.~\ref{fig3}, where the second and third excited states are
depicted in natural scale versus the coupling constant $g$. This
figure contains a fragment of the near-threshold area in larger
scale in order to make the third level visible, since it looks
almost like a dot and cannot be distinguished from the
two-particle threshold within the main figure. Note that the
largest binding energy of each of the Efimov levels (minus the
energy of the two-particle threshold) happens at the constants
rather far from the critical concentration point of the spectrum
(see Fig.~\ref{fig2} and Fig.~\ref{fig3}), though the infinite
number of levels appears only at the limiting point $g_{cr}$. The
three-particle levels cross the two-particle threshold (with
increase in $g$) at very small angles at the points in
Fig.~\ref{fig3}, where the solid lines of these levels become the
dashed line of the two-particle threshold. After crossing the
two-particle threshold, the three-particle levels may exist over
the threshold as virtual states (see~\cite{R16}). It is worth to
note that, with the essential enlargement of the coupling constant
$g$ up to $g \ge g_{cr}^{(2)}=16.3$, the second excited level
appears below the two-particle threshold, the third one does at $g
\ge g_{cr}^{(3)}= 30.8$, the fourth does at $g \ge
g_{cr}^{(4)}=32.55$, the fifth does at $g \ge g_{cr}^{(5)}=
50.89$, and so on. Their binding energies increase with the
coupling constant. The angular coefficients determining the angle
between the curve of the $n$-th level and the two-particle energy
threshold (near the critical point of the corresponding level
appearance),
\begin{equation}
E_n  - E_{(2)}  \cong \tilde k_n \left( {g - g_{cr}^{(n)} }
\right)\, , \label{E13}
\end{equation}
are as follows: $\tilde k_2 \cong - 0.0045$, $\tilde k_3 \cong -
0.1$, $\tilde k_4 \cong - 0.01$, and $\tilde k_5 \cong - 0.03$.
The fact that, for the third and  fourth levels, $g_{cr}^{(3)}$
and $g_{cr}^{(4)}$ are close one to another, as well as that the
angular coefficients $\tilde k_3$ and $\tilde k_4$ are of
different order of magnitude, testifies to a different nature of
these levels.

\section{Structure Functions of the Efimov States}
The wave functions of the three-particle ground and excited states
calculated within the precise variational approach enable us to
calculate directly such structure functions of the system as
density distributions, formfactors, pair correlation functions,
momentum distributions, etc. It is a rather simple problem to
calculate any such an average since the wave functions of each
state have the simple form (\ref{E3}) of a superposition of
Gaussian functions with the already known linear ($D_k$) and
nonlinear ($a_k$, $b_k$, $c_k$) parameters.

\begin{figure}[ht]
\centering \includegraphics[width=4.5in]{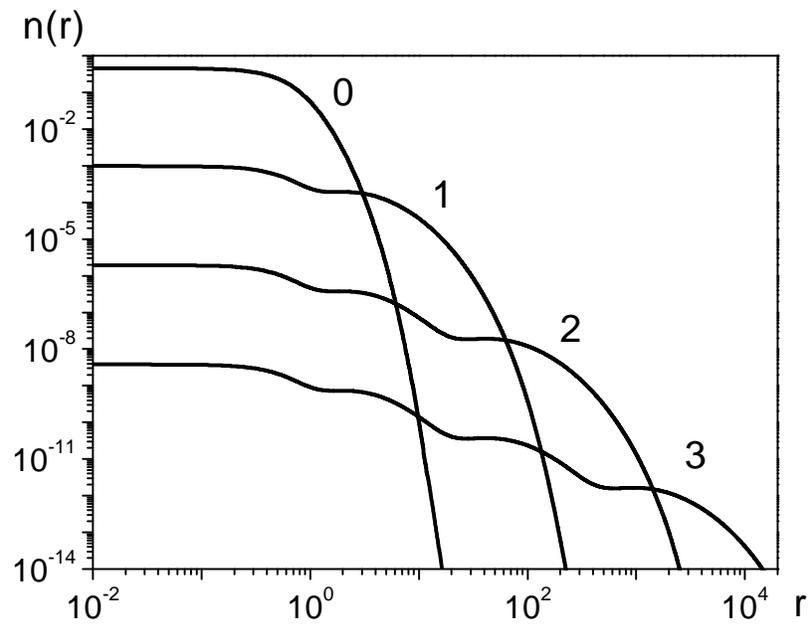}
\caption{One-particle density distributions for the ground ($n=0$)
and excited ($n=1,2,3$) states.} \label{fig4}
\end{figure}

Fig.~\ref{fig4} presents the one-particle density distributions,
\begin{equation}
n(r) = \left\langle \frac{1}{3} \sum\limits_{i = 1}^3 \delta
\left({\bf r} - \left({\bf r}_i - {\bf R}_{c.m.} \right)\right)
\right\rangle \, ,\label{E14}
\end{equation}
for the ground and excited states at the critical two-particle
coupling constant $g_{cr}$. The density distributions (\ref{E14})
are shown in logarithmic scale on both axes. This is made, on the
one hand, in order that the typical ``halo'' structure appearing
for different states at distances of different scales can be shown
in the same figure. On the other hand, on the chosen logarithmic
scale, the density distributions reveal clearly the almost
periodic dependence of the wave functions of Efimov states on the
global radius logarithm \cite{R1} $\sim \sin \left( \left| s_0
\right|\ln \left( R\sqrt {m\left| E \right|/\hbar ^2 } \right) +
\Delta \right)$. Moreover, this asymptotic behavior is valid from
the distances of about the radius of forces (in our case, of about
1) to the distances of about $\hbar \left( m \left| E
\right|\right)^{-1/2}$. The less the binding energy of an excited
state, the longer is the extension of this asymptotic behavior.
The density distribution of the first excited state is seen to
change sharply its behavior between the short and long distances.
The long-range ``halo'' is situated around the more condensed
central core similar to the density distribution of the ground
state. The density distribution of the second excited state
changes its behavior two times having a ``halo'' in the form of
two concentric spherical layers put one into another. They are of
essentially different radii and densities and have the central
part of the same type as in the case of the first excited state.
The density distribution of the third excited state changes its
behavior three times (three concentric layers around the central
core), and so on. The ratio of the radii of neighboring layers is
about $ \sqrt {\lambda _E } \cong 22.7$. Since the density
distributions are normalized by $\int n(r)d{\bf r} = 1$, they
decrease at short distances when the number of the excited state
increases, because the size of the system is growing.

\begin{figure}[ht]
\centering \includegraphics[width=4.5in]{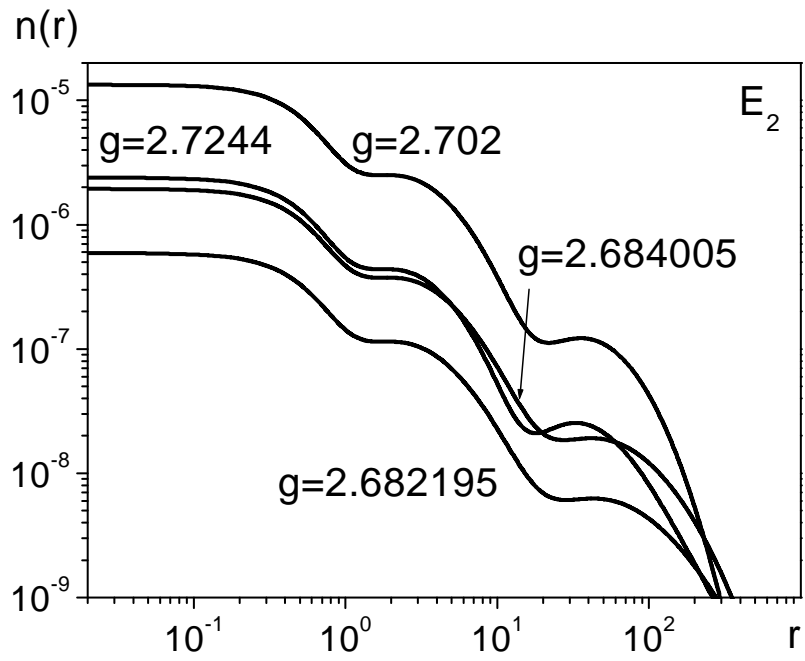}
\caption{Density distribution profiles with several different
coupling constants $g$ for the second excited state.} \label{fig5}
\end{figure}

The fact that the ratio of the sizes of neighboring states of the
system is about $\sqrt {\lambda _E }$ is a consequence of the
universal model-independent properties of Efimov states and is
confirmed once more by calculations of the r.m.s. radii (see
Table \ref{tab1} with the calculated $R_{rms}$ at the critical
value $g_{cr}$). Note that the smallest value of $R_{rms}$ for a
given Efimov level is achieved at the constant $g$ corresponding
to the largest binding energy, rather far to the right from the
critical point $g_{cr}$ (see Fig.~\ref{fig2}). When the coupling
constant approaches the value, where the Efimov levels appear at
the threshold ($E \to 0$), the calculated mean square radius
\begin{equation}
R_{rms}  \equiv \left\langle {r^2 } \right\rangle ^{1/2}  =\left(
\int {r^2 n(r)d{\bf r}}\right)^{1/2} \label{E15}
\end{equation}
reveals the behavior
\begin{equation}
\left\langle {r^2 } \right\rangle ^{1/2}  = \frac{c_1 }{\sqrt E }
+ c_2  + \dots \, .\label{E16}
\end{equation}
That is, the radius goes to infinity, in accordance with the
general physical principles, as $\left\langle {r^2 } \right\rangle
^{1/2} \sim (E)^{-1/2}$ at $E \to 0$ (we recall that the energy
level near the threshold depends linearly on the coupling constant
according to (\ref{E12})). It is important to notice that the
growth of the r.m.s. radii at $E \to 0$ is connected mainly with
the extension of the region where the asymptotic behavior of the
wave function $\sim \sin \left( \left| s_0 \right|\ln \left(
R\sqrt {m\left| E \right|/\hbar ^2 } \right) + \Delta \right)$ is
valid, i.e., up to the distances of order $\hbar \left( m \left| E
\right|\right)^{-1/2}$. At the same time, the wave function at
short and intermediate distances does not change (except for the
normalization factor depending on the total size of the system).
As a result, in all the intervals of $g$ where a given Efimov
level exists, the behavior of the wave function and, in
particular, the density distribution is not changed at short and
intermediate distances (but only the normalization does), which is
confirmed by Fig.~\ref{fig5}. This figure shows the density
distributions of the second excited state at several different
values of the coupling constant, from the point $g = 2.682195$
near the level appearance to the point $g_{right}^{(2)} = 2.7244$
of its disappearance. Other typical constants chosen by us are as
follows: the two-particle critical constant $g_{cr}$ (the point of
the Efimov spectrum concentration) and the constant $g = 2.702$,
where the energy of the second excited state (minus the
two-particle threshold energy) is the greatest.

\begin{figure}[ht]
\centering \includegraphics[width=4.5in]{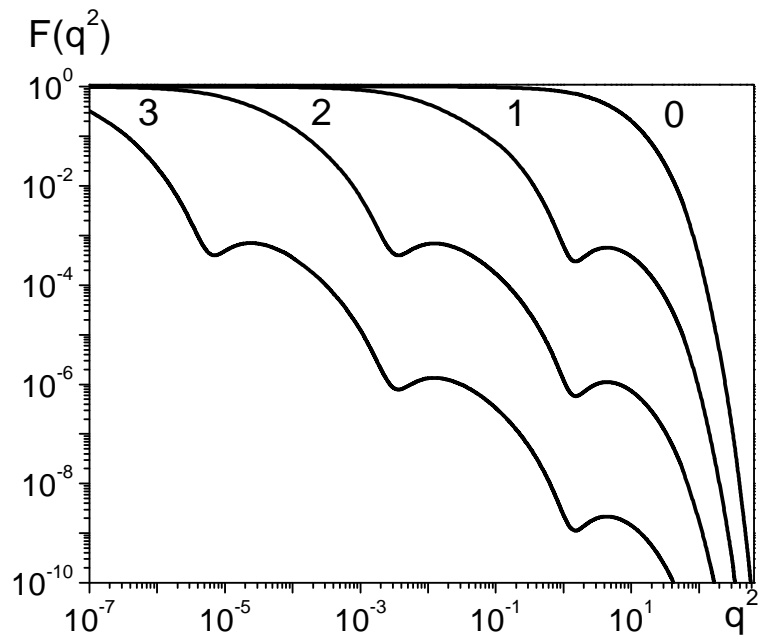}
\caption{Formfactors for the ground ($n=0$) and three excited ($n
=1,2,3$) states.} \label{fig6}
\end{figure}

Specific sharp changes in the density profiles near the distances,
where the corresponding wave functions pass through zero, manifest
themselves in formfactors,
\begin{equation}
F(q) = \int \exp ( - i{\bf q} \cdot {\bf r})n(r)d{\bf r}\, ,
\label{E17}
\end{equation}
in the form of specific dips of finite depth. Fig.~\ref{fig6}
shows the formfactors of the ground state (no dips), the first
(one dip), second (two dips), and third (three dips) excited
states. A logarithmic scale commonly used for formfactors on the
vertical axis is accomplished with the logarithmic scale in the
squared momentum transfer in order to show clearly the almost
periodic repetition of the dips connected with regularities in the
``halo'' structure of the density distributions of excited states.
Since the greater distances in a density distribution correspond
to a less momentum transfer in the corresponding formfactor and
vice versa, a new dip in the formfactor of the next excited state
appears at a less momentum transfer, whereas the previous dips
remain practically at their positions.

\begin{figure}[ht]
\centering \includegraphics[width=4.5in]{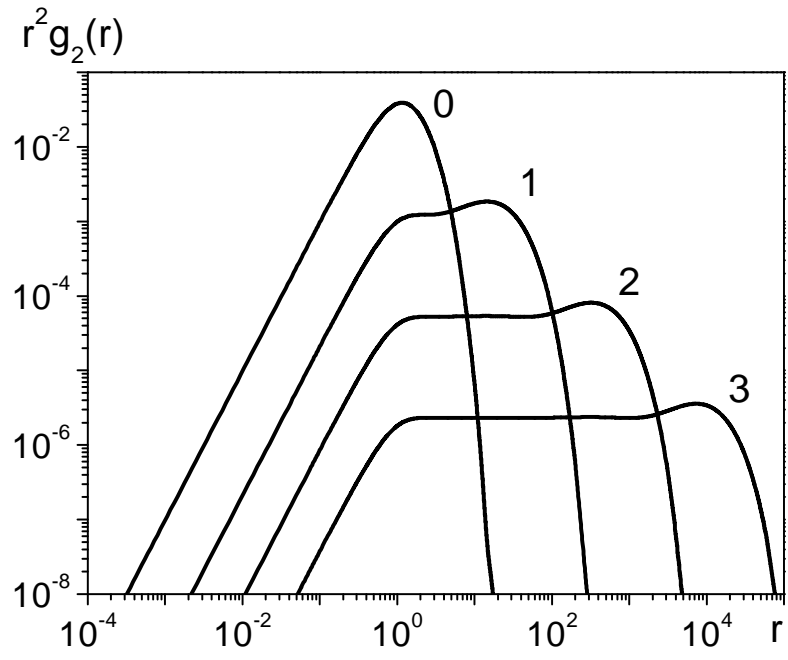}
\caption{Pair correlation functions for the ground ($n=0$) and
three excited ($n=1,2,3$) states.} \label{fig7}
\end{figure}

Fig.~\ref{fig7} shows the pair correlation functions
\begin{equation}
g_2 (r) = \left\langle \frac{1}{3} \sum\limits_{i
> j = 1}^3 \delta \left({\bf r} - \left({\bf r}_i  - {\bf r}_j
\right)\right)
\right\rangle \label{E18}
\end{equation}
of the ground state and three excited states. We multiply the pair
correlation functions by $r^2$ in order to show clearly the
following interesting effect. With increase in the number of the
excited state, we observe the spreading of the area, where the
pair correlation function is proportional to the reciprocal
squared distance between a pair of particles. That is, the
two-particle subsystem in the system of three particles has the
density distribution very similar to the squared two-particle wave
function. This gives rise to the effective long-range interaction
\cite{R1,R2,R3} $\sim - R^{-2}$ in the system of three particles
in the region from distances of order of the radius of forces
$R\sim r_0$ to the distances of about $R\sim \hbar \left(m\left| E
\right|\right)^{-1/2}$, where the wave function starts its
exponential decrease due to a finite binding energy.

\begin{figure}[ht]
\centering \includegraphics[width=4.5in]{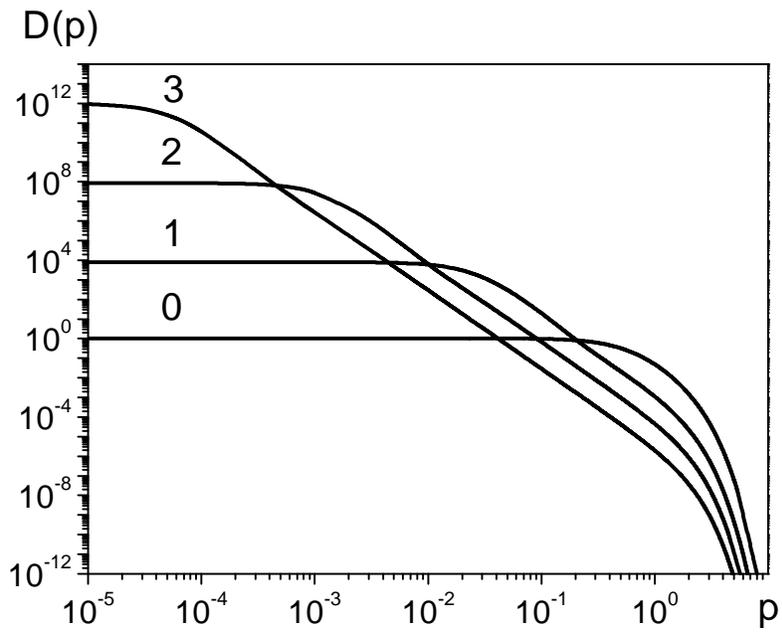}
\caption{Momentum distributions for the ground ($n=0$) and three
excited ($n=1,2,3$) states.} \label{fig8}
\end{figure}

The momentum distributions
\begin{equation}
D(p) = \left\langle \frac{1}{3} \sum\limits_{i = 1}^3 \delta
\left({\bf p} - {\bf p}_i \right) \right\rangle \label{E19}
\end{equation}
for the ground and three excited states are presented in
Fig.~\ref{fig8}. It is found that the momentum distributions
vanish at approximately the same limiting momentum $p_0$ (in our
case, $10^0 < p_0  < 10^1$). This is due to the fact that the
largest momenta in the considered three-particle system are
typical of the shortest distances in the coordinate representation
(of order of the radius of forces). There, the wave function has
the sharpest behavior, but it is practically unchangeable under
growing the number of an excited state. With increase in the
number of a state, the wave function extends to larger and larger
distances almost without change in the central part (see, for
illustration, Fig.~\ref{fig4} depicting the density
distributions). Note that, with increase in the number $n$ of the
state, the momentum distribution $D(p)$ increases essentially at
small momenta proportionally to $\left( {\lambda _E ^{3/2} }
\right)^n$. Indeed, since the size of the system increases by
$\sqrt {\lambda _E }$ times for each of the next excited states
and its volume increases by $\left( {\lambda _E } \right)^{3/2}$
times, the region of momenta where the momentum distribution is
mainly concentrated decreases at the same extent. Taking into
account $\int D(p)d{\bf p} = 1$, we get that the momentum
distribution density grows by $\left( {\lambda _E } \right)^{3/2}
\approx 1.17 \times 10^4$ times at small $p$ with every increment
of the state number. Fig.~\ref{fig8} also demonstrates the
extension (towards zero momentum) of the area of the power
dependence of the momentum distribution $D(p)\sim p^{-4}$, which
looks like a linear decreasing function on the logarithmic scale.
Such a behavior is connected with the fact that the greater the
number $n$ of the state, the greater is the role of the kinetic
energy entering the Green's function in the formation of the power
dependence of the wave function in the momentum representation.
Really, in the formal operator representation of the
Schr\"{o}dinger equation solution
\begin{equation}
\left. \left| {\Psi _n }\right. \right\rangle  = ( H_0  - E_n )^{
- 1}  V\left. \left| {\Psi _n }\right. \right\rangle \, ,
\label{E20}
\end{equation}
the energy $E_n \to 0$ with increase in $n$. In this case, instead
of the Green's function, we have only the kinetic energy in the
denominator. The expression $V\left. \left| {\Psi _n }\right.
\right\rangle$ varies slightly, because it is nonzero only within
the region of the short-range interaction, where the wave function
is changed insignificantly at growing $n$. After the integration
with the $\delta$ - function, the squared modulus of the wave
function in $D(p)$ leads to the value in the denominator which is
proportional to the squared kinetic energy of one particle, by
giving rise to $D(p)\sim p^{-4}$.

\section{Conclusions}
We have developed a variational approach with the use of the
Gaussian basis and precise optimization schemes and have studied
the spectrum and wave functions of three particles with Gaussian
pairwise interaction near the critical coupling constant, where
the Efimov effect takes place. A special technique is developed
for preparing the initial configurations of highly excited
near-threshold weakly bounded energy states using both the
deformation of the initial Hamiltonian (by changing the
interaction or the mass ratio) and the method of evolution in the
coupling constant at a fixed (in particular, zero) energy. For
several calculated levels from the Efimov series, it is shown that
they have an essential asymmetry with respect to the critical
point and the largest binding energy (minus the threshold energy)
significantly far to the right from the critical coupling constant
which is the point of the spectrum concentration. The scale
invariance is confirmed for the Efimov levels starting from the
second excited state of three particles.

For the first time, we have carried out the study of the specific
behavior of the density distributions, formfactors, pair
correlation functions, and momentum distributions for the ground
and three excited states of the Efimov series. The details of the
density distributions are found, and the halo-type structure is
revealed for the excited states. It is shown that the
corresponding formfactors have some dips of finite depth which are
arranged periodically in the momentum transfer on the logarithmic
scale. The behavior of pair correlation functions and momentum
distribution profiles is found, and the main conclusions following
from the asymptotic formulae for the Efimov spectrum are
confirmed. The main specific properties of the structure functions
of three particles within the area of parameters where the Efimov
effect reveals itself would take place with other attractive
interaction potentials as well.

The precise calculations of the spectrum and the wave functions of
highly excited near-threshold levels confirm the high accuracy of
the variational method in the Gaussian representation combined
with the efficient technique for the maximum optimization of the
basis.

\clearpage
\begin{table}
\caption{ Characteristics of three-particle energy
levels in the Efimov effect region. In the second and fourth
columns, the corresponding values of the reciprocal scattering
length $a^{-1}$ are given for each value of the coupling constant
$g$.}
\begin{tabular}{|l|c|c|c|c|c|}
\hline $n$&$g_{left}^{(n)}$&$k_n  = \partial E_n / \partial
g$&$g_{right}^{(n)}$&$E_n$&$R_{rms}$\\
&$(a^{-1})$&&$(a^{-1})$&$(a^{-1}=0)$&$(a^{-1}=0)$\\ \hline
0&2.13079&-0.147&-&-0.23845&1.21
\\ &(-0.22853)&&&&\\ \hline 1&2.64396&-4.72$\times
10^{-3}$&-&-4.514$\times 10^{-4}$&2.25$\times 10^{1}$ \\
&(-1.35046$\times 10^{-2}$)&&&&\\ \hline 2&2.68215&-2.07$\times
10^{-4}$&2.7244&-8.76$\times 10^{-7}$&5.04$\times 10^{2}$\\
&(-6.1717$\times 10^{-4}$)&&(1.32485$\times 10^{-2}$)&&\\ \hline
3&2.68393&-1.38$\times 10^{-5}$&2.68567&-1.7$\times
10^{-9}$&1.14$\times 10^{4}$\\ &(-2.48262$\times
10^{-5}$)&&(5.53501$\times 10^{-4}$)&&\\ \hline
\end{tabular}
\label{tab1}
\end{table}
\clearpage
\clearpage
\clearpage
\clearpage
\clearpage
\clearpage
\clearpage
\clearpage

\end{document}